\documentclass[preprint,nonacm]{acmart}

 \copyrightyear{2026}
 \acmYear{2026}
 \setcopyright{cc}
\acmDOI{10.1145/3805689.3812401}

\usepackage{longtable}
\usepackage{booktabs}
\usepackage{array}
\usepackage{tabularx}
\usepackage[table]{xcolor}
\usepackage{pdflscape}
\usepackage{longtable}
\newcommand{\heat}[2]{\cellcolor{blue!#1}{#2}}
\renewcommand{\arraystretch}{1.1}
\begin{document}

\title{From Cradle to Cloud: A Life Cycle Review of AI’s Environmental Footprint}

\author{Katherine Lambert}
\affiliation{%
  \institution{University of Toronto}
  \city{Toronto}
  \country{Canada}
}

\author{Sasha Luccioni}
\affiliation{%
  \institution{Hugging Face, McGill University}
  \city{Montreal}
  \country{Canada}}

\renewcommand{\shortauthors}{Lambert and Luccioni}

\begin{abstract}

The rapid growth in the deployment and scale of modern artificial intelligence (AI) systems has intensified concerns regarding their environmental impacts, yet we still lack a comprehensive view of where and how these impacts arise across the AI life cycle. In order to shed more light on this question, we conduct a structured, comprehensive literature review of scientific papers and technical reports that examine different aspects of AI's environmental footprint. Using an eight-stage life cycle framework, spanning hardware manufacturing, infrastructure construction, data gathering and preprocessing, model experimentation, training, post-training adaptation, deployment, inference, and end-of-life, we systematically map which stages are covered, the  metrics reported at each stage, and the methodological choices made. We then draw conclusions about the information we gathered, finding that although life cycle language is increasingly common in discussions of “green” or “sustainable” AI, its definition remains unclear -- while some studies focus solely on model training and inference, others encompass broader measurements such as data collection, infrastructure, and embodied emissions. We also find that reporting practices rely predominantly on CO$_2$e estimates derived from coarse proxies, with limited attention dedicated to water usage, materials manufacturing, and multi-impact life cycle assessment, making it difficult to compare and aggregate true results. Building on these findings, we propose measurement and reporting approaches to support more comprehensive, comparable and policy-relevant assessments of AI's environmental impacts. 
\end{abstract}




\maketitle


\section{Introduction} \label{sec:introduction}

    The recent development and deployment of artificial intelligence (AI) models has been accompanied by far-reaching impacts on society, from labor markets~\cite{eloundou2024gpts}, to criminal sentencing procedures~\cite{angwin2022machine} and education~\cite{blodgett2021risksaifoundationmodels}. Over the past decades, extensive bodies of work have examined the way these models are developed and how design and deployment choices translate into intended and unintended consequences for different groups and populations~\cite{buolamwini2018gender,crawford2021atlas, bender2021dangers}. Furthermore the environmental impacts of AI's widespread deployment are profound, notably due to the heavy consumption of natural resources~\cite{falk2025flopsfootprintsresourcecost,falk2025carboncradletograveenvironmentalimpacts}, energy and water required for powering and cooling the data centers needed to train and serve models~\cite{luccioni2022estimating,Luccioni_2024,li2023making,lei2025water}, as well as the greenhouse gases (GHGs) emitted across all stages of the AI supply chain~\cite{luccioni2023counting,dodge2022measuring,strubell2019energy}. Despite these impacts and their consequences, the magnitude and extent of these effects remain largely undisclosed by model developers~\cite{luccioni2025misinformation, luccioni2023counting}, making it difficult to conduct rigorous assessments and audits of system impacts.
    
Our study addresses this gap through a structured literature review of 61 works on AI's environmental impacts published between 2019 and 2025, following a multi-stage methodology described in Section~\ref{sec_methodology}. We analyze the literature through an eight-stage AI life cycle lens (which we define in Section~\ref{sec_lifecycle}), which enables us to compare where environmental impacts are currently measured, which stages remain underexplored, what metrics are reported at each stage, and how measurement practices differ across studies.

We present these results in Section~\ref{sec:results}, highlighting several common focal points, namely:
\begin{enumerate}
    \item Overall, the existing literature remains concentrated on operational stages, especially training and inference, which leaves important stages such as dataset creation, post-training adaptation, and end-of-life substantially under-examined. 
    \item Most studies emphasize energy use and carbon emissions, while broader impacts such as water use, material extraction, toxicity, and embodied infrastructure burdens are far less consistently reported.
    \item Reporting practices remain heterogeneous in scope, metrics, and measurement methodology, which limits comparability across studies and weakens the basis for governance, auditing, and standardized disclosure.
\end{enumerate}

We then use these observations to identify methodological gaps and propose clearer reporting directions for future work, which we discuss in Section~\ref{sec:discussion}.

\section{Related Work} 
\label{sec:related-work}

    Research on the environmental impacts of AI has been an increasingly prolific field of scholarship, with the number of articles written on the topic increasing dramatically in recent years (as we show in Figure~\ref{fig:paper_growth}). Much of this work, which we survey in our literature review, is dedicated to carrying out empirical studies regarding the environmental impacts of AI models and approaches -- studies like Strubell et al.~\cite{strubell2019energy}, Luccioni et al~\cite{luccioni2024power,luccioni2022estimating} and Henderson et al~\cite{henderson2020towards}, which carry out model pre-training or fine-tuning and estimate the amount of energy used and GHGs emitted. There are also numerous papers that aim to propose approaches and strategies for mitigating AI's environmental impacts, e.g., Schwartz et al.~\cite{schwartz2020green}, who introduce the concept of ``Green AI'', Dodge et al.~\cite{dodge2022measuring}, who look at different approaches to minimize the emissions of training AI models on cloud-based infrastructure, and Luccioni et al.~\cite{luccioni2023counting}, who survey the factors that influence the final carbon footprint of AI model training, such as the energy source that was used, as well as model size and modality.  

    While questions of AI ethics and sustainability have mostly been treated separately, recent scholarship in the fairness, accountability, and transparency community has begun to analyze how the two topics intersect and can be considered in tandem~\cite{varoquaux2025hype,luccioni2025bridging,van2021sustainable}. For instance, Bender, Gebru et al.~\cite{bender2021dangers} hone in on large language models (LLMs) as a type of AI model that can have particularly negative consequences both for the environment as well as in terms of ethics and fairness, especially given the ubiquity of their deployment. In a similar vein, Luccioni et al.~\cite{luccioni2025efficiency} look at the trade-offs between efficiency and rebound effects, as well as indirect impacts in terms of space, time and scale, arguing that different criteria have to be taken into account to make informed decisions about AI's environmental impacts.  Varoquaux et al.~\cite{varoquaux2025hype} go a step further, examining broader trends in AI, by which the explosive growth in model size and compute requirements translate into an trajectory that is unsustainable from several different perspectives, from economic to environmental. All of these papers call for more transparency and formal disclosure in terms of the costs of model training and inference, without which it is impossible to make informed decisions and perform analyses on their impacts.

In terms of literature reviews of environmental impact research, two prior works are especially relevant to our study. Verdecchia et al.~\cite{verdecchia2023systematic} survey what they call ``Green AI'' and find that observational studies focus primarily on training, with only limited attention dedicated to inference. Similarly, Barbierato et al.~\cite{barbierato2024toward} analyze sustainable versus unsustainable AI research trends across model families and application areas, arguing that sustainability should be considered alongside performance. We build on these reviews but contribute a different organizing perspective: rather than grouping studies by topic or type alone, we compare them through an explicit life cycle lens that enables cross-stage analysis of coverage, metrics, and measurement practices. This perspective is central to our contribution: rather than treating environmental reporting as a question limited to training or inference, we examine how existing work distributes attention across upstream, operational, and downstream stages of AI systems. We present the methodology of our literature review in more detail  below.
 
\section{Methodology} \label{sec_methodology}

    In order to be as comprehensive as possible in our approach, we conducted a structured literature review of work published between 2019 (the date when Strubell et al's first paper regarding AI's environmental impacts was published~\cite{strubell2019energy}) and 2025, to examine how the environmental impacts of AI were measured and reported across this period of time. Our analysis focused on three key dimensions -- life cycle coverage, reported metrics, and methodological choices -- and our literature review proceeded in three steps: (1) defining an analytical life cycle framework, (2) collecting and screening relevant studies, and (3) extracting comparable information from the final corpus.
    
    We structured our investigation around the following research questions:
    
    \begin{enumerate}
        \item To what extent has the environmental impact of AI systems been studied in the existing literature?
        \item Which stages of the AI life cycle are most commonly analyzed, and which remain underexplored?
        \item What types of environmental metrics are reported, and how are they measured?
        \item How do methodologies differ across approaches in terms of tools, estimation techniques, and infrastructure focus?
    \end{enumerate}
    
    These questions guided the search, screening, and categorization of the papers we surveyed, as well as the analysis of reporting practices, tools, models, and life cycle coverage. Our ultimate aim was to provide a comprehensive assessment of current approaches, which can inform the development of both more holistic environmental evaluation methods, as well as more principles policy approaches. 

    \subsection{Analytical Life Cycle Framework} \label{sec_lifecycle}

We used an eight-stage AI life cycle framework as an analytical device to examine how environmental impacts are studied in the existing literature on the topic. It should be noted that we do not present this framework as a universal or strictly sequential model of AI development; rather, it is a structured decomposition for environmental assessment that makes it possible to compare studies with different scopes and system boundaries. This framing is particularly useful because while we found that the term ``AI life cycle'' has clear, agreed-upon boundaries in sustainability and policy discourse~\cite{heijungs2010life,sala2016life}, in AI it is used inconsistently, with some papers using it to refer only to training and inference, and others including additional stages. 

The 8 stage framework described below was derived by comparing life cycle concepts used in prior environmental assessment and AI governance literature, including LCA-oriented work and research papers on AI environmental assessment~\cite{luccioni2023estimating,de2022artificial,ligozat2022unraveling}, broader Information and Communications Technology (ICT) life cycle models~\cite{arushanyan2014lessons,guldbrandsson2012opportunities}, ISO 14040/14044-oriented life cycle assessment foundations~\cite{finkbeiner2006new}, and higher-level AI life cycle frameworks such as those proposed by the OECD and the ISO. Across these sources, the common underlying idea is that AI systems can be analyzed as a set of stages associated with different activities, resources, and environmental impacts. For instance, the OECD AI life cycle model adopts a more model-centric approach without considering, e.g., hardware manufacturing or data center heating, ventilation and air conditioning (HVAC) systems~\cite{oecd2025}, whereas the ISO/IEC 5338:2023 standard does not specify the metrics that should be declared at each stage of the life cycle~\cite{iso5338}. Contrary to these, our framework follows that shared logic while introducing more granularity for stages that are often collapsed in prior work, especially hardware and infrastructure impacts, post-training adaptation, and end-of-life.

The following criteria guided the delineation of stages:

        \begin{itemize}
            \item \textbf{Functionally ordered:} The stages are presented in an analytically useful order from upstream production to downstream use and retirement,  mirroring the LCA logic of ``cradle-to-grave'' assessment. Acknowledging that AI development is often iterative and that some stages may recur or overlap, this framework is intended as an \emph{analytical decomposition} for environmental assessment rather than a universal or strictly sequential model of AI development.
            
            \item \textbf{Distinct in impact:} Each stage represents a qualitatively different set of environmental processes and burdens -- including embodied emissions, operational energy demand, water consumption, and e-waste generation -- allowing for differentiated assessment of material and energy flows across the life cycle.
            
            \item \textbf{Empirically tractable:} The stages are defined such that their impacts can be reasonably estimated or measured using available or emerging data sources and tools. Where direct measurement is infeasible, proxy metrics (e.g., FLOPs, GPU-hours, total number model parameters) and extrapolation methods from policy or forecasting studies are considered valid indicators.
            
            \item \textbf{Appropriate level of granularity:} The framework balances analytical precision with interpretability, maintaining enough resolution to distinguish major contributors to environmental impact while remaining generalizable across diverse study types -- including empirical measurements, conceptual frameworks, and trend analyses.
        \end{itemize}

        \begin{figure}[h!]
          \centering
          \includegraphics[width=\textwidth]{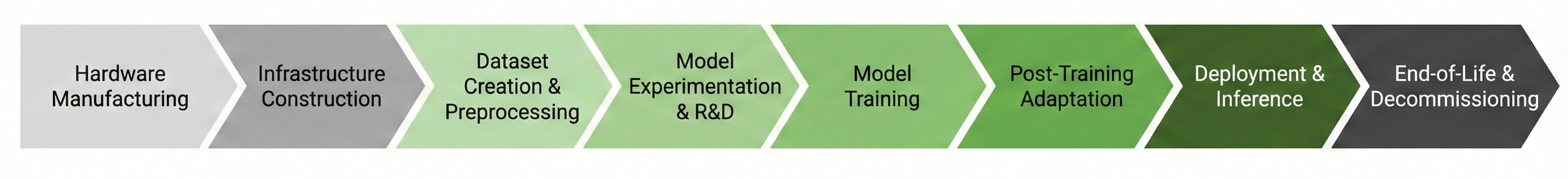}
          \caption{The 8 AI life cycle stages identified in our analysis.}
          \label{fig:ai_stages}
        \end{figure}
        
        Guided by these principles and informed by the patterns observed in the reviewed works, we categorize the AI life cycle into eight distinct yet interlinked stages, also represented in Figure~\ref{fig:ai_stages}:
        \paragraph{1. Hardware Manufacturing}
        
        This stage encompasses the extraction of raw materials, fabrication of computing chips (e.g., CPUs, GPUs, TPUs), and the assembly of servers and other hardware used in AI. It includes embodied emissions and resource use associated with production, such as carbon, water consumption, and mineral depletion.
        
        \paragraph{2. Infrastructure Construction} 
        
        This includes the construction and operation of physical infrastructure required to host and power AI workloads: data centers, construction and assembly of networking hardware, as well as cooling and backup systems. It also captures environmental impacts and embodied emissions associated with cement, steel, land use change, and the transportation and construction of equipment.
        
        \paragraph{3. Dataset Creation and Preprocessing}
        
        This stage refers to all activities involved in acquiring, curating, cleaning, augmenting, filtering, or synthesizing training data. It includes the environmental footprint of Web crawling, storage, and preprocessing. In some newer models (e.g., LLM distillation workflows) and models such as DeepSeek v3~\cite{liu2024deepseek}, it also includes synthetic data generation, which is an emerging energy hotspot in model development pipelines.
        
        \paragraph{4. Model Experimentation and R\&D}
        
        This stage includes the iterative experimentation processes that precede final model training, such as hyperparameter tuning, ablation studies, and architectural exploration, which can be a substantial fraction of the total model emissions and compute, as referenced in recent LLM case studies~\cite{morrison2025holistically,luccioni2023estimating}.
        
        \paragraph{5. Model Training}
        
        The stage involves the final training run of a model on the full dataset, typically using GPUs, TPUs, or other accelerators. 
        
        \paragraph{6. Post-Training Adaptation}
        
        This stage encompasses fine-tuning, distillation, instruction tuning, quantization, and other post-training modifications that adapt the base model to specific tasks or domains. It is especially relevant in transfer learning and deployment scenarios, where base (or `foundation'~\cite{bommasani2021opportunities}) models are often repurposed or adapted to specific deployment contexts.
        
        \paragraph{7. Deployment and Inference}
        
        This stage includes the environmental impact of deploying the model in production and running inference workloads over time. Depending on usage volume, inference contributes a large portion of a model's lifetime energy consumption, particularly for AI models deployed at scale~\cite{wu2022sustainable}.
        
        \paragraph{8. End-of-Life and Decommissioning}
        
        The final stage covers the retirement of hardware and infrastructure, including e-waste disposal, recycling, or repurposing. This phase is crucial for a full cradle-to-grave analysis, and it involves consideration of toxic materials, recyclability of GPUs and server components, and the environmental burdens of safe disposal.
        
        \medskip
        
        Together, these stages provide a comprehensive framework for  environmental life cycle assessment of AI systems. This structured life cycle categorization forms the analytical backbone of our study, which we describe below, and we propose it as a basis for future standardized assessments of AI’s environmental footprint.

    \subsection{Search Strategy and Paper Collection}

        Searches were run in Fall of 2025 on Semantic Scholar, arXiv, ACM Digital Library, and Google Scholar. We used combinations of AI-related terms (“artificial intelligence”, “AI”, “machine learning”, “deep learning”, “large language model”, "LLM") \footnote{The AI-related terms were treated interchangeably for search purposes to increase findings as much of the relevant literature, especially earlier work, is framed as machine learning, or LLM-related work, while studying systems that fall within the broader AI ecosystem.} with environmental-assessment terms (“environmental impact”, “carbon emissions”, “energy consumption”, “water use”, “life cycle assessment”, "life cycle", “environmental footprint”, “sustainability”, “reporting”, “measurement"). We initially retrieved 82 records, screened titles and abstracts for relevance, and then conducted full-text review of the remaining papers against the inclusion criteria above. Duplicates, papers with no substantive AI/ML focus, papers that mentioned sustainability only in passing, and works that did not quantify, estimate, or methodologically discuss environmental impacts of AI systems were excluded. We supplemented database search with backward citation chaining and author-based exploration: for highly relevant papers, we examined their reference lists and checked whether their authors had published additional work on AI environmental measurement or life cycle assessment. This process yielded a final corpus of 61 works, papers and reports spanning 2019–2025.         
        
        Given that there is no single publication venue for this subject, we included peer-reviewed articles, preprints, whitepapers, and blog posts that reported empirical data, life cycle estimates, or relevant methodology. Eligible studies quantified AI’s environmental footprint -- energy, emissions, water, or life cycle stages -- from hardware to deployment. Table~\ref{tab:venue_counts} shows that the literature is dispersed across venues, with no central publication source -- this fragmentation makes discovery and aggregation of papers more difficult.

        \begin{table}[h!]
            \centering
            \begin{tabular}{lcl}
                \toprule
                \textbf{Venue type} & \textbf{Count} & \textbf{Example venues / notes} \\
                \midrule
                Journals / magazines & 26 (42\%) & \textit{Comms. of the ACM}, \textit{Sustainable Computing}, \textit{ACM SIGEnergy}, etc. \\
                Conferences / workshops & 14 (23\%) & ACL, FAccT, NAACL, IEEE HPEC, etc. \\
                Preprint (arXiv) & 18 (29\%) & \\
                Other (industry, blogs) & 3 (5\%) & Includes reports and blog-style analyses \\
                \bottomrule
            \end{tabular}
            \caption{Publication venues for the 61 works in our analysis.}
            \label{tab:venue_counts}
        \end{table}
        
\begin{figure}[h!]
          \centering
          \includegraphics[width=0.6\textwidth]{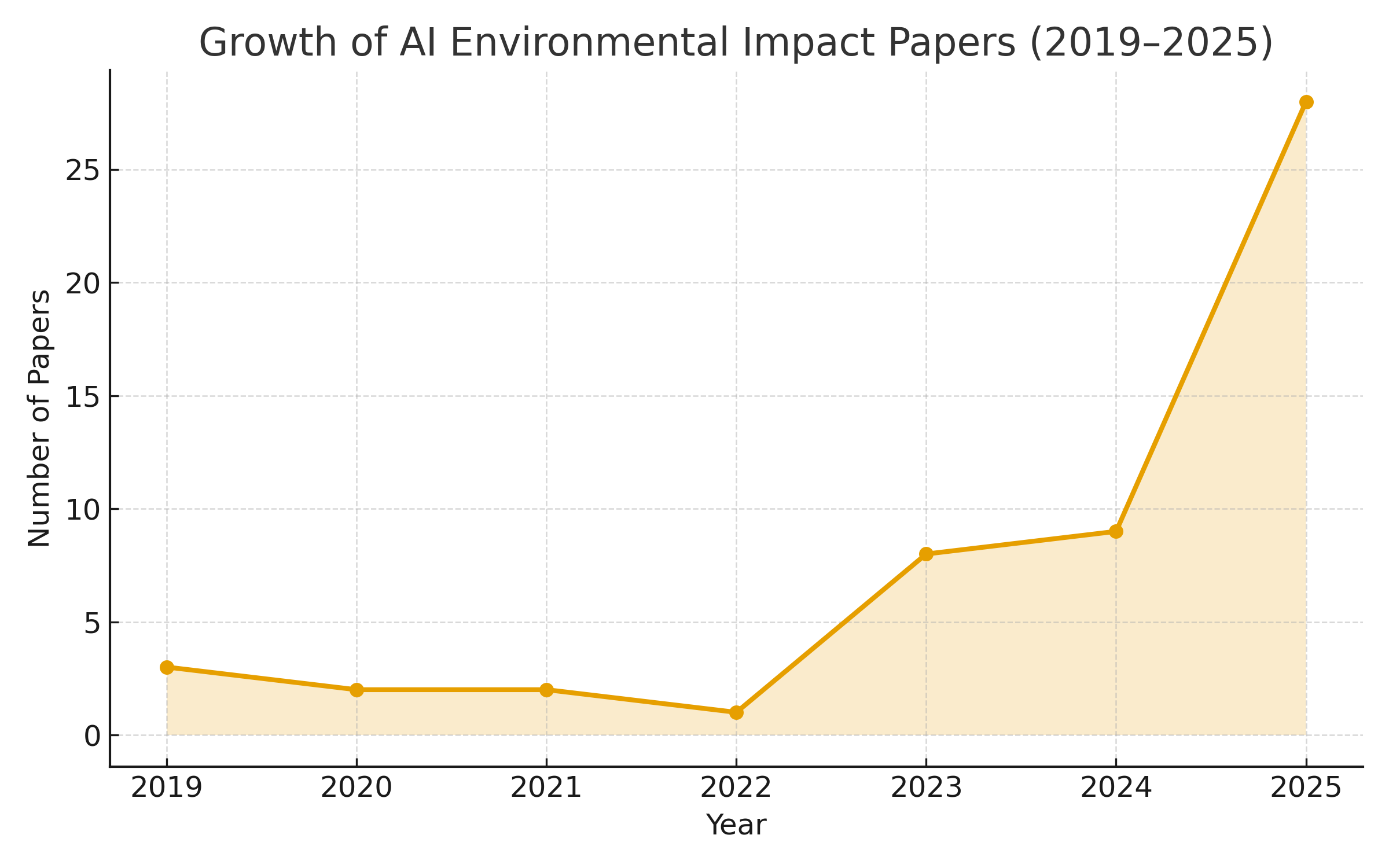}
          \caption{Number of papers included in the review, by publication year.}
          \label{fig:paper_growth}
        \end{figure}

        The earliest life cycle-focused studies appeared in 2019, emphasizing training-related energy and emissions (e.g., ~\citet{strubell2019energy, henderson2020towards}), when AI-specific environmental reporting became more visible and systematic; earlier literature did not yield many papers that reported AI-specific environmental metrics in a manner comparable to the studies within our corpus. In comparison to the earlier works, more recent papers (from 2025) increasingly address underexplored stages such as inference, hardware manufacturing, and infrastructure. This is shown in Figure~\ref{fig:paper_growth}, which illustrates rising interest in this topic, with a noticeable shift toward broader life cycle coverage in 2025, as well as a drastic increase in the overall number of papers between 2023 and 2025. 
        
    \subsection{Data Extraction and Categorization}
        
        All 61 papers were reviewed in detail and catalogued in a structured spreadsheet supporting both quantitative and qualitative analysis (which we provide as a \href{https://docs.google.com/spreadsheets/d/e/2PACX-1vT99H-f3ACoQ4VEARjC1x2aJG4hfHbKxVb8XytazibOQd48fpWccFgwC4kRLTlV6KcXA6lvd7uNPvDm/pubhtml}{complement} to our study). For each article, we recorded core metadata, including title, year, venue, summary, and thematic tags indicating paper type (e.g., empirical measurement, theoretical analysis, policy discussion, framework) -- see Table~\ref{app:paper-table} in the Supplementary Materials for the full list of papers. To enable a more life cycle-focused comparison, each paper was labeled according to the AI life cycle stage(s) it addressed, based on those listed in Section~\ref{sec_lifecycle}. We used multi-label tagging: papers were assigned to all life cycle stages that were substantively analyzed, rather than only to a single primary stage. Stage assignment was based on the paper's stated research questions, empirical focus, and the environmental impacts actually reported, rather than on brief mentions of other stages.
        
        In addition to this information, we extracted methodological details for each paper: whether energy or emissions were measured directly, estimated from hardware specifications, modeled theoretically, or computed using tools such as CodeCarbon~\cite{schmidt2021codecarbon}, CarbonTracker~\cite{anthony2020carbontracker}, or Experiment Impact Tracker~\cite{henderson2020towards}. We documented all reported metrics (i.e., energy, CO$_2$e, water, cost, FLOPs, GPU-hours), the hardware used (e.g., A100/V100/H100 GPUs, TPUs), and the types of models analyzed (e.g., GPT-3, BERT, T5, or model-agnostic studies). For papers spanning multiple activities, categorization was based on the portions of the life cycle that were actually measured, estimated, or discussed in methodological detail. Finally, the spreadsheet captured each paper's key results, takeaways, reported limitations, and identified gaps, which we use as the basis of the analysis presented below.

        We do recognize that our review has several methodological limitations. First, because the literature is fragmented across journals, conferences, preprints, whitepapers, and industry publications, exhaustive retrieval is difficult. Second, because we included non-peer-reviewed sources when they reported otherwise unavailable empirical estimates, source quality and reporting granularity are heterogeneous. Third, life cycle-stage assignment required interpretive judgment, especially for papers spanning multiple phases or using broad terms such as ``training'' to cover several life cycle stages. Finally, despite broad search, citation chaining, and author-based exploration, some relevant studies may have been missed. Even with these limitations, this study provides, to our knowledge, the most comprehensive literature review of AI's environmental impacts to date, illuminating both established findings and critical research gaps. We present the results of our analysis in the following section.

\section{Results} \label{sec:results}

    In the sections below, we report how the literature is distributed across life cycle stages, and then summarize the environmental metrics and measurement approaches used across the corpus. The interpretation of these findings and their implications are deferred to Section~\ref{sec:discussion}.


    \subsection{Life Cycle Coverage}

        Our review, as reported in Table \ref{tab:lifecycle_coverage_table}, finds that existing work is heavily concentrated on model training and, increasingly, deployment and inference, while upstream (hardware manufacturing, infrastructure construction, dataset creation) and downstream stages (experimentation, post-training adaptation, end-of-life) remain sparsely quantified or only treated qualitatively. This uneven coverage limits current LCAs to partial, operation-centric views rather than veritable cradle-to-grave assessments.
   \begin{table}[h!]
        \centering
        \begin{tabular}{l c}
            \toprule
            \textbf{AI life cycle stage} & \textbf{Number of papers} \\
            \midrule
            Hardware Manufacturing              & 20 \\
            Infrastructure Construction         & 25 \\
            Dataset Creation \& Preprocessing   & 5  \\
            Model Experimentation \& R\&D       & 8  \\
            Model Training                      & 31 \\
            Post-Training Adaptation            & 0  \\
            Deployment \& Inference             & 36 \\
            End-of-Life \& Decommissioning      & 8  \\
            \bottomrule
        \end{tabular}
        \caption{Count for number of reviewed papers across each AI life cycle stage. N.B. that many papers refer to multiple life cycle stages, which is why the total number of papers is more than 61.}
        \label{tab:lifecycle_coverage_table}
    \end{table}

       \subsubsection{Hardware Manufacturing}
                
            The environmental impacts of hardware manufacturing are markedly underrepresented in AI sustainability research. Across the 61 papers in our corpus, only 20 explicitly cover hardware manufacturing or hardware life-cycle impacts. Most studies instead adopt an operational boundary focused on electricity use during training or inference, sometimes with coarse device-level emission factors that implicitly assume CPU-style hardware \cite{luccioni2023estimating,luccioni2025misinformation}. Among the subset that does address manufacturing, there is a clear move toward full life-cycle assessments of GPUs, TPUs, and related infrastructure, combining process-based models, vendor data, and physical teardowns to obtain device-specific footprints \cite{schneider2025life,falk2025more}. In parallel, an increasing number of vendors are releasing product-level carbon footprints; for example, NVIDIA’s cradle-to-gate assessment of the HGX H100 GPU baseboard reports about 1,312 kg CO$_2$e per unit, a substantial amount of emissions that further reinforces the importance of manufacturing in footprint analyses \cite{nvidiaHGXH100PCF}.
    
            These emerging assessments converge on several qualitative findings. First, embodied emissions from chip fabrication, memory, and supporting infrastructure can be of the same order of magnitude as operational emissions over typical device lifetimes, challenging evaluations that treat hardware as impact-free capital \cite{schneider2025life}. Second, when impacts are broken down across categories beyond climate change, for instance, mineral depletion, ecotoxicity, or human health, manufacturing processes and rare-metal supply chains often dominate, revealing trade-offs that are invisible in carbon-only metrics \cite{luccioni2025misinformation,falk2025carboncradletograveenvironmentalimpacts}. Finally, the literature highlights substantial methodological and data gaps: LCAs frequently depend on proprietary process information or single-vendor product carbon footprints, and there is still no systematic, comparative assessment of embodied impacts across GPU generations or architectures. Recent framework and tooling efforts call for more transparent, standardized models for including hardware and infrastructure in AI LCAs, but practical guidance on how to do so at scale remains limited \cite{plociennik2025life,simon2024boaviztapi}. As a result, hardware manufacturing is widely acknowledged as a critical blind spot, and the lack of comparable embodied-impact estimates currently limits our ability to evaluate hardware-level efficiency trends or justify design choices on environmental grounds.

        \subsubsection{Infrastructure Construction}
                
            Infrastructure construction and operation impacts are referenced in 25 of the 61 papers in our corpus, but most do so only indirectly, via aggregate facility-level overheads, sometimes combined with regional grid carbon intensity \cite{lacoste2019quantifying,patterson2021carbon}. In these studies, infrastructure largely appears as a multiplicative factor on IT energy, rather than as a life cycle stage with its own embodied emissions from materials such as concrete, steel, mechanical and electrical systems, or specialized cooling and power distribution equipment \cite{alissa2025using,lei2025water}. A smaller subset introduces additional facility metrics, such as Water Usage Effectiveness (WUE~\footnote{Water Usage Effectiveness (WUE) is a data center sustainability metric defined as total site water use (for cooling, humidity control, etc.) divided by IT energy use, typically reported in L/kWh or m$^3$/MWh.}) and water footprint estimates, to capture location and design-specific trade-offs between carbon, energy, and water use for AI workloads, particularly for large-scale inference \cite{jegham2025hungry,chien2023reducing,li2025making}. 
            
            Only a few assessments attempt to explicitly attribute embodied impacts to infrastructure construction, either by allocating portions of data center LCAs to AI workloads (which is difficult to do given the lack of granular data regarding what percentage of data center workloads can be directly attributed to AI) or by modeling cooling and auxiliary systems at component level \cite{alissa2025using,lei2025water,schneider2025life}. Even in these cases, infrastructure-related emissions are often reported coarsely, and the one-time impacts of greenfield buildout, grid connection upgrades, and end-of-life treatment remain sparsely quantified. However, as hyperscale data center buildout accelerates, this gap limits our ability to understand how rapidly expanding data center fleets reshape life cycle burdens and to compare infrastructure choices  --  such as cooling strategies, siting, or reuse of existing facilities  --   on a consistent resource cost and environmental basis~\cite{hintemann2022cloud,d2024carbon,lei2025water,herrera2025sustainable}.

        \subsubsection{Dataset Creation and Preprocessing}
            The environmental footprint of dataset acquisition, curation, and preprocessing is one of the least empirically studied stage in AI LCAs. Only 5 of the 61 papers in our corpus ($\sim$8\%) explicitly model impacts from dataset gathering, acquisition, or preprocessing. This stage is often acknowledged in high-level life cycle diagrams, but very rarely quantified: only a handful of studies attempt to account for energy, water, or resource use for data pipelines, and existing tracking tools overwhelmingly focus on training and inference rather than upstream data work. Conceptual LCA frameworks argue that data-related activities should fall within scope by mapping AI tasks to the ICT equipment used for data collection, storage, and processing, and then assessing those devices over manufacturing, use, and end-of-life stages~\cite{ligozat2022unraveling, plociennik2025life}. In practice, however, most empirical assessments either omit data processing entirely or subsume it into a generic “overhead,” leaving upstream stages such as data acquisition, large-scale web crawling, and complex preprocessing (tokenization, deduplication, filtering) effectively unreported.

            Emerging evidence from web-scale dataset and data-infrastructure projects suggests that these activities can carry non-trivial operational footprints, yet they are rarely integrated into downstream model LCAs. For example, constructing multimodal corpora such as LAION-400M/5B and the DataComp candidate pool involves petabyte-scale crawling, long-term storage over many TB-months, and repeated CLIP-based filtering, but these works are typically described only in terms of dataset size or model quality, with no accompanying environmental accounting~\cite{Schuhmann2021LAION400MOD,Schuhmann2022LAION5BAO,Gadre2023DataCompIS}. Recent work on data filtering further shows that curation itself can be a significant and scale-dependent consumer of compute~\cite{Goyal2024ScalingLF}, while reporting guidelines and existing carbon-accounting studies still seldom link these costs back into model-level life cycle assessments~\cite{henderson2020towards}. Likewise, the lack of transparency around dataset creation can become an increasingly glaring oversight given the advent of synthetic data generation and its role in the performance of models such as DeepSeek R1~\cite{deepseekai2025deepseekr1incentivizingreasoningcapability}, without a clear indication of the cost of these improvements and the relative comparison of this step to other ones in the AI model life cycle. 
            
            In principle, practitioners could meter CPU/GPU energy for preprocessing jobs, attribute network transfer energy by the quantity of data moved and track storage as TB-hours with region-specific power and water factors, integrating these into attributional LCA inventories \cite{ligozat2022unraveling}. A comprehensive LCA of AI systems should therefore elevate dataset creation and preprocessing from a nominal box in a diagram to a first-class, metered phase, with reproducible disclosure of at least preprocessing compute, data movement, and storage, alongside any embodied impacts of dedicated data infrastructure.

        \subsubsection{Model Experimentation and R\&D}
    
            The environmental impact of model experimentation and research iterations -- hyperparameter tuning, architecture search, ablations, and exploratory runs -- is widely recognized as a major driver of AI’s total compute use, but is rarely measured systematically. In our corpus, only 8 of the 61 papers (about 13.1\%) explicitly treat experimentation or hyperparameter search as a distinct life cycle stage, and only a few papers provide separate estimates for the incremental energy or emissions of development runs beyond training a single final configuration. However, early empirical analyses of NLP models showed that neural architecture search can increase emissions by orders of magnitude relative to training a single configuration, with individual development campaigns reaching hundreds of tons of CO$_2$e \cite{strubell2019energy}. Conceptual work on “Green AI” formalizes this by treating the number of experimental trials (hyperparameter experiments) $H$ as a key multiplier of development cost, and argues that efficiency metrics and “price tags” for compute should be reported alongside accuracy \cite{schwartz2020green}. In our corpus, however, only a small minority of studies attempt to quantify energy use for R\&D phases; most report the footprint of one “final” training run and omit the experiments that led to it. Where development is quantified, reports for large language models increasingly indicate that the cumulative compute used for experimentation can rival or exceed that of the final training run~\cite{zhang2022opt}, and a recent third-party analysis of OpenAI's compute suggests that most of it went to experimentation, not training or inference~\cite{epoch2025openaicomputespend}. This suggests  that ignoring R\&D can underestimate total life cycle impacts by several orders of magnitude.
            
            A growing ecosystem of measurement tools makes it technically feasible to account for these costs at the level of individual experiments. Frameworks such as Carbontracker and CodeCarbon enable per-run logging of energy, carbon intensity, and hardware usage \cite{anthony2020carbontracker,schmidt2021codecarbon}, and recent datasets instrument thousands of training runs with wattmeter-level detail. For example, the BUTTER-E dataset adds measured energy consumption to over 60,000 deep learning experiments, revealing that architectural and hyperparameter choices can produce superlinear increases in energy for diminishing accuracy gains \cite{tripp2024measuring}. Together, these results point to experimentation as a large but often hidden component of AI’s environmental footprint. Yet, few AI LCAs aggregate energy across all trials, and almost none report basic statistics such as the number of runs, total GPU-hours, or search strategies used. Establishing norms for reporting project-level R\&D energy -- rather than only the final training job -- thus remains a key opportunity for making life cycle assessments more realistic and for incentivizing more efficient experimental practices.

        \subsubsection{Model Training}
                
            The model training stage is one of the most extensively quantified components of AI’s environmental life cycle. In our corpus, just over half of the papers report energy use or associated greenhouse gas emissions for training runs, often making training the primary focus of their empirical analysis. In typical practice, authors translate runtime into energy by multiplying GPU-hours by device Thermal Design Potential (TDP~\footnote{Thermal Design Power (TDP) is the maximum heat/power a processor is designed to dissipate under typical maximum load, used as a design target for cooling and power supply capacity.}), then multiplying the result by regional grid carbon intensity~\footnote{Some of these even use average US carbon intensity instead of a region-specific value to avoid revealing data center locations.} and, sometimes, a facility-level overhead factor such as PUE \cite{lacoste2019quantifying,patterson2022carbon}. Public calculators and tooling (e.g., MLCO2 Impact~\cite{lacoste2019quantifying}) enable this accounting by asking users for hardware type, runtime, and cloud region, returning approximate CO$_2$e estimates based on assumed power and regional emission factors. These approaches have revealed strong sensitivity to siting and system design: training the same model on a coal-heavy grid versus a low-carbon grid can change emissions substantially, and differences in data center efficiency (PUE $\approx$ 1.1–1.8 versus the implicit assumption of PUE = 1.0) further modulate total system energy \cite{lacoste2019quantifying}. 
            
            Across the corpus, several recurring insights emerge. First, training footprints vary widely as a function of hardware choice, software and algorithmic efficiency, and workload placement, with optimized configurations often delivering substantial reductions in energy and emissions for a fixed model quality \cite{patterson2022carbon}. Second, reported figures frequently omit important components of training-related energy, such as failed or aborted runs, warm-up and debugging phases, or infrastructure overheads beyond the accelerator itself, leading to systematic underestimation of life cycle impacts \cite{strubell2019energy,schwartz2020green}. Third, the emphasis is put on models of a bigger size and from the textual modality (i.e., LLMs), with other modalities and sizes being under-represented. Finally, documentation practices remain inconsistent: many papers report only high-level runtime and hardware labels, without specifying GPU-hours, exact device models, grid emission factors, or whether measurements include CPU and RAM energy, cooling and power distribution. As with other life cycle stages, the literature therefore points to a need for standardized reporting conventions for training -- covering, at minimum, GPU-hours, average power, grid and PUE assumptions, and scope (device-only vs. full facility) -- to enable meaningful comparison across models and to connect model-level benchmarks to system-level environmental outcomes \cite{lacoste2019quantifying,plociennik2025life}.

        \subsubsection{Post-training Adaptation}        
            Post-training adaptation (e.g., fine-tuning, reinforcement learning, distillation, pruning, or quantization) has become a central stage in contemporary AI workflows as practice shifts toward reusing large pre-trained models, yet it remains largely invisible in environmental accounting. In our corpus, none of the 61 papers explicitly model adaptation as a separate life cycle stage; at most, a small handful mention these techniques in passing or subsume them under a single undifferentiated “training” phase. As a result, the incremental energy and emissions associated with adapting foundation models to downstream tasks are almost never reported, despite their growing share of real-world AI workloads. Fine-tuning is frequently characterized as a small fraction of pre-training cost, depending on model size and task \cite{strubell2019energy, henderson2020towards, luccioni2023estimating}. However, when popular foundation models are adapted by many downstream users across organizations and domains, the aggregate energy for fine-tuning, supervised adaptation, and reinforcement-learning-based alignment can collectively rival or exceed that of a single pre-training run, even though detailed carbon accounting for these activities is rarely disclosed \cite{luccioni2025misinformation}. 
            
            Beyond standard fine-tuning, a growing set of post-training approaches -- including parameter-efficient adaptation, continual and domain-specific updates, distillation, and quantization -- further complicate the life cycle picture. Distillation and post-training quantization can substantially reduce inference-time energy per query, but the energy and emissions invested in creating distilled or compressed models are almost never reported or weighed against downstream savings in a life cycle framework \cite{yang2024double}. Likewise, continual adaptation in deployed systems (for example, reinforcement learning from human feedback, or coding models that continue to improve based on verified rewards) introduces recurring post-training overhead that typically remains unaccounted for. Across the literature we reviewed, few studies specify whether reported “training” footprints include any of these adaptation steps, and almost none break out their contributions separately. As development practices converge on model reuse and iterative adaptation, future LCAs should treat post-training as a distinct life cycle stage, with explicit reporting of compute, energy, and emissions, to avoid systematically underestimating its impacts \cite{luccioni2025misinformation}.

        \subsubsection{Model Deployment \& Inference}
        
            Across the life cycle literature, model deployment (also called model inference or serving) is now widely recognized as major drivers of energy and carbon, especially for high-volume generative services. In our corpus, 36 of the 61 papers (59\%) report deployment- or inference-stage impacts, compared to 31 (50\%) that quantify training. Earlier work overwhelmingly emphasized training and often treated inference as a negligible usage of energy or source of emissions, but this balance has shifted: deployment-stage assessments now appear slightly more often than training-focused assessments, reflecting a rapid change from its previous characterization as ``understudied''~\cite{luccioni2024power}. Collectively, these studies conclude that for always-on, large-scale systems, the use phase can dominate the one-off training run, with estimates ranging from ``inference comparable to training'' to ``inference providing the large majority of lifetime energy use'', and even modelling scenarios where annual serving emissions exceed training by more than an order of magnitude~\cite{chien2023reducing,jegham2025hungry,yang2024double}. They also highlight how sensitive per-query energy is to deployment choices -- including batch size and concurrency, sequence length, and software stack -- and document sharply diminishing accuracy gains relative to increases in inference energy as models scale~\cite{yang2024double}.
            
            A growing body of systems work examines how to mitigate this footprint in practice, typically combining hardware-efficient serving with strategies such as aggressive batching, dynamic load balancing, and carbon- or energy-aware routing and scheduling across heterogeneous data centres\cite{chien2023reducing,jegham2025hungry}. However, most such studies are conducted in controlled environments or as one-off case studies, and only a small number of industrial reports provide full-stack measurements of production LLM services, including idle capacity, non-accelerator components, cooling, and water~\cite{elsworth2025measuring}. Public analyses continue to stress limited transparency and partial reporting around production inference workloads, even as per-query metrics for individual models become more common, leaving important gaps around edge deployments, user-side energy use, long-lived model maintenance, and standardised reporting practices for inference-stage emissions.\cite{plociennik2025life}

        \subsubsection{End-of-Life (EoL) \& Decommissioning}
                
            End-of-life is one of the least covered life cycle stages in our review: only 8 of the 61 papers in our corpus (13.1\%) mention hardware or system disposal, recycling, or material recovery. This mirrors a broader ICT pattern where analysis typically stops at the operational phase, despite LCA guidance to include all stages~\cite{luccioni2025misinformation}. The few cradle-to-grave assessments of AI accelerators and data center hardware that do exist partition impacts into embodied, use-phase, and EoL contributions, and converge on a common message: while end-of-life usually accounts for only a small fraction of total climate impacts (often a few percent), it can be much more significant for categories such as mineral and metal depletion, human toxicity, and ecotoxicity~\cite{schneider2025life,falk2025more}. These studies highlight that disposal of high-density electronics concentrates critical and hazardous materials in relatively small masses, creating non-trivial risks and missed recovery opportunities that are largely invisible in carbon-only accounting.
            
            Beyond AI-specific work, ICT and “responsible digital” frameworks increasingly advocate circular economy strategies -- extending hardware lifetimes, cascading devices into less demanding roles, refurbishing accelerators, and improving recycling yields -- as levers to reduce e-waste and embodied impacts~\cite{luccioni2023estimating}. However, AI life cycle assessments rarely quantify such scenarios or report concrete hardware fates, recovery rates, or data-security constraints at decommissioning. EoL processes thus remain a critical blind spot: their direct CO$_2$ contribution is modest, but their indirect effects via resource loss, hazardous waste, and material scarcity are substantial. To produce credible, ISO-aligned assessments at AI scale, future work will need to systematically integrate decommissioning pathways, region-specific recycling assumptions, and transparent reporting on hardware lifetimes, reuse, and recovery outcomes.


        \subsection{Metrics Reported} \label{subsec_metrics}
        
            As shown in Table~\ref{tab:lifecycle_metric_coverage_heatmap}, energy consumption is the most consistently reported metric across the literature. Most studies in our corpus quantify electricity use (typically in kWh or joules) and convert it to carbon emissions (CO$_2$e) using an emissions factor. These two metrics, energy and carbon, dominate the field and function as the de facto standard indicators of environmental impact. It should be noted that such conversions can often rely on broad assumptions, such as average grid intensity (which is not necessarily representative of the energy mixes of specific data centers, which may have on site generators) and Power Usage Effectiveness (PUE) values, commonly assumed to be $\sim$1.5–1.8, which may not reflect the actual data center or hardware context, thereby limiting comparability and precision across studies.                     

            \begin{table*}[h!]
                \centering
                \scriptsize
                \setlength{\tabcolsep}{2pt}
                \renewcommand{\arraystretch}{1.1}

                \begin{tabular}{lccccccccc}
                    \toprule
                    & \textbf{Hardware} & \textbf{\begin{tabular}[c]{@{}c@{}}Infra\\  Construction\end{tabular}} &
                 \textbf{\begin{tabular}[c]{@{}c@{}}Dataset\\  Creation\end{tabular}} & \textbf{\begin{tabular}[c]{@{}c@{}}Model\\  R\&D\end{tabular}} &
                      \textbf{Training} & \textbf{\begin{tabular}[c]{@{}c@{}}Post-train.\\   Adaptation\end{tabular}} & \textbf{\begin{tabular}[c]{@{}c@{}}Deployment/\\   Inference\end{tabular}} &
                 \textbf{\begin{tabular}[c]{@{}c@{}}EoL/\\   Decomm.\end{tabular}} &
                      \textbf{Overall} \\
                    \midrule
                    
                    \textbf{Energy (kWh)} &
                    \heat{11}{5} &
                    \heat{15}{6} &
                    \heat{3}{1} &
                    \heat{8}{4} &
                    \heat{24}{8} &
                    \heat{0}{0} &
                    \heat{30}{11} &
                    \heat{3}{1} &
                    \heat{64}{36} \\
                    
                    \textbf{GHG / CO$_2$e} &
                    \heat{11}{4} &
                    \heat{14}{5} &
                    \heat{2}{1} &
                    \heat{6}{2} &
                    \heat{20}{7} &
                    \heat{0}{0} &
                    \heat{18}{6} &
                    \heat{4}{2} &
                    \heat{55}{27} \\
                    
                    \textbf{Water Use} &
                    \heat{3}{1} &
                    \heat{6}{2} &
                    \heat{1}{1} &
                    \heat{1}{1} &
                    \heat{6}{2} &
                    \heat{0}{0} &
                    \heat{6}{2} &
                    \heat{2}{1} &
                    \heat{25}{10} \\
                    
                    \textbf{Multi-impact LCA} &
                    \heat{4}{2} &
                    \heat{2}{1} &
                    \heat{1}{1} &
                    \heat{0}{0} &
                    \heat{3}{1} &
                    \heat{0}{0} &
                    \heat{3}{1} &
                    \heat{3}{1} &
                    \heat{16}{7} \\
                    
                    \textbf{Embodied HW impacts} &
                    \heat{5}{2} &
                    \heat{6}{2} &
                    \heat{0}{0} &
                    \heat{1}{1} &
                    \heat{3}{1} &
                    \heat{0}{0} &
                    \heat{4}{2} &
                    \heat{2}{1} &
                    \heat{21}{9} \\
                    
                    \textbf{Compute (FLOPs / GPU-h)} &
                    \heat{6}{2} &
                    \heat{10}{4} &
                    \heat{1}{1} &
                    \heat{7}{3} &
                    \heat{14}{5} &
                    \heat{0}{0} &
                    \heat{13}{5} &
                    \heat{3}{1} &
                    \heat{34}{21} \\
                    
                    \textbf{Cost / \$} &
                    \heat{0}{0} &
                    \heat{2}{1} &
                    \heat{0}{0} &
                    \heat{4}{2} &
                    \heat{5}{2} &
                    \heat{0}{0} &
                    \heat{2}{1} &
                    \heat{0}{0} &
                    \heat{13}{6} \\
                    
                    \bottomrule
                \end{tabular}
                \caption{Metric-by-life cycle coverage heatmap: number of papers reporting each metric at each AI life cycle stage.}   \label{tab:lifecycle_metric_coverage_heatmap}
                {\footnotesize
                    \textit{Note:} Rows correspond to reported environmental \& proxy metrics; columns correspond to AI life cycle
                    stages. Cell values show the number of papers that explicitly report a given metric at that life cycle stage; color
                    intensity encodes relative frequency across all cells.
                }
            \end{table*}
            This focus on reporting energy use and GHG emissions leaves other impact dimensions such as water, toxicity, and resource depletion rarely  quantified. For instance, fewer than 20\% of papers report water usage, which has only recently begun to receive attention from the AI community, despite the rising quantity of water needed to cool AI data centers ~\cite{lei2025water, herrera2025sustainable}. However, Li et al.~\cite{li2025making} showed that training GPT-3 may have directly evaporated around 700{,}000 liters of water in Microsoft’s U.S. data centers, and highlight a fundamental trade-off: carbon-efficient daytime scheduling can conflict with water-efficient nighttime cooling. Similarly, life-cycle studies of AI hardware, such as Google’s TPU fleet~\cite{schneider2025life} and the recent analysis of Nvidia A100 GPUs~\cite{falk2025more}, find that manufacturing and end-of-life stages dominate human toxicity and mineral depletion, even when operational carbon remains the largest climate contributor. 
            
            Furthermore, only a small subset of papers log hardware‑level performance data (e.g., memory usage, GPU utilization), often when using tools such as CarbonTracker, CodeCarbon, or 
            Experiment Impact Tracker. FLOPs and model parameters are occasionally reported, mainly as proxies for computational work or for normalizing results. Across studies, measurement methodology varies widely, from direct power monitoring (e.g., NVIDIA NVML, Intel RAPL) to software-based estimators and analytical approximations using FLOPs, runtime, or TDP. This variation leads to uneven reporting and a strong bias toward energy and carbon metrics, with comparatively little coverage of water, materials, or other life cycle burdens.  Different papers use different units, scopes, and assumptions, which hampers comparisons and progress. For example, one might report “GPU hours” without specifying if that refers to a single GPU or a whole cluster-hour; others might include an industry average PUE of 1.8 while another assumes a much lower one that better reflects a specific data center. Framework papers and tools (e.g., Ligozat et al.~\cite{ligozat2022unraveling}, Berthelot et al.~\cite{berthelot2024estimating}, Luccioni et al.~\cite{luccioni2025misinformation}, Boavizta~\cite{simon2024boaviztapi}) therefore call for multi-criteria, ISO-aligned assessment and greater methodological transparency, and there is a general consensus in recent literature that methodological inconsistency is a problem --  we echo this in Section~\ref{subsec_proposals}.
            

\section{Discussion and Proposals} \label{sec:discussion}

While carrying out an analysis of the state-of-the-art research on AI's environmental impact was the primary goal of our study, we also intend for these insights to be useful to a broader public, beyond AI researchers and scholars. In the current section, we put forward proposals for improving the state of AI environmental impact assessment, as well as the implications for policymaking. 

    \subsection{Proposals for Improving AI Environmental Impact Measurement and Reporting} \label{subsec_proposals}

    Our literature review yields four main recommendations for standardizing and strengthening environmental impact assessment in AI. These recommendations are intended to guide AI researchers and developers in reporting their environmental impacts more comprehensively:

        \paragraph{Improve life cycle comprehensiveness} Measuring the true environmental cost of AI requires benchmarking impacts across all stages of the AI life cycle. We propose using the 8 stages defined in Section~\ref{sec_lifecycle} as the basis for future environmental impact analyses. While all steps are not relevant for all kinds of studies, using them to identify \emph{which} stages of the life cycle are analyzed (e.g., "We study the environmental impacts of synthetic data generation, as part of the Data Creation and Preprocessing life cycle stage") could help others use the takeaways to inform subsequent work. 
        
        \paragraph{Use standard metrics} A practical LCA framework for AI should standardize what and how to measure environmental impacts. This includes defining the units and scope (e.g., report kWh and $CO_{2e}$ for at least training and inference, use consistent emissions factors, include precise PUE and carbon intensity values). The lack of standardization noted by many authors could be addressed by a framework that tailors existing LCA standards (ISO 14040/44) to AI. For example, the framework might prescribe reporting of energy per training run, energy per inference, carbon emissions (with location-based factors),  along with context like hardware type and utilization. A more comprehensive approach should push beyond carbon to include water footprint metrics (WUE) and possibly other impacts (if data allows), to guide sustainability-informed decision making. It might also include a template to report per-query water consumption alongside energy and carbon for inference services – something virtually unheard of in current papers, but made plausible by recent research.
         
        \paragraph{Use tools for direct measurement} While using proxies such as GPU hours and average grid carbon intensities can contribute towards comparing orders of magnitudes of total energy usage or emissions, it does little to allow for more granular comparisons between different hardware types, usage intensities, and geographical variation. Using tools such as CodeCarbon or CarbonTracker enables direct measurements of energy usage as well as up-to-date carbon intensity figures, which results in more fundamentally-sound comparisons. The success of these tools suggests that providing accessible instrumentation encourages researchers to report energy, and their integration into conferences such as \href{https://cvpr.thecvf.com/Conferences/2026/ComputeReporting}{CVPR} indicates an interest in the community for this kind of reporting and longitudinal tracking. Providing the raw measurements from these tools -- similar to what was done by the BLOOM model carbon footprint paper authors~\cite{luccioni2022estimating} -- could also help fill data gaps and enable further analyses (e.g., component carbon footprints, standardized hardware specifications). However, it should also be noted that these tools are not a complete solution: their estimates depend on hardware detection, system boundaries, and assumptions about carbon intensity and infrastructure overhead, so they should be reported alongside their underlying assumptions rather than treated as full ground truth.

        \paragraph{Translate empirical results into actionable insights.}  Translating technical conclusions and analyses into more accessible terms can make these results actionable for non-technical audiences. This has already been done by some researchers -- for instance, Luccioni et al.'s~\cite{luccioni2024power} study on AI inference ends with high-level takeaways that are put in non-technical terms; similarly, Dodge et al.~\cite{dodge2022measuring} provide carbon and energy equivalents that allow for comparisons with different fields and applications.  We are currently at a crucial point in AI regulation when regulators are looking to researchers for experimental insights to inform and guide their policymaking~\cite{hacker2023regulating}; by making results interpretable to non-technical stakeholders, researchers can better inform evidence-based regulation and contribute to aligning environmental objectives with organizational decision-making.

  \subsection{Policy Implications}

Our literature review also has direct implications on the development of AI-specific policymaking frameworks and regulations. For instance, it can help improve the comprehensiveness and coverage of environmental impact metrics and disclosures -- this can, in turn, contribute towards encouraging policymakers to develop additional reporting requirements. These requirements can include detailed environmental reports disaggregated by model, usage patterns, and physical infrastructure, and enforcement mechanisms such as regular comprehensive environmental reports. These can then be used as incentives for rewarding environmentally-minded model developers and providers, for instance by making access to government funding and contracts limited to those who disclose the necessary metrics.

Further, the adoption of more comprehensive environmental disclosure metrics is a critical prerequisite for developing international standards, which would facilitate the integration of AI-related emissions  into broader climate policy instruments, such as carbon pricing schemes and corporate sustainability reporting, while also enabling the development of benchmarking systems or environmental labels for AI services -- such as the AI Energy Score project~\cite{luccioni2024light}, which is currently voluntary and limited to open-source models, but can be converted into a more formal mechanism if integrated into regulatory frameworks.

Third, the use of direct measurement tools introduces the possibility of more robust compliance and auditing mechanisms. Tools such as CodeCarbon and CarbonTracker lower the barrier to accurate reporting and could form the basis of standardized monitoring requirements. Policymakers may require their use in regulated contexts, alongside disclosure of underlying assumptions and raw measurement data to support third-party verification. Over time, this could enable continuous monitoring of large-scale AI systems and foster the emergence of independent auditing ecosystems, strengthening the credibility and enforceability of environmental disclosures.

Finally, and in complement to the fourth suggestion made in Section~\ref{subsec_proposals}, working hand-in-glove with AI researchers to translate their findings into actionable and accessible insights is an essential contribution towards bridging the gap between technical research and policymaking, and will require efforts made on both sides. For instance,  if an LCA reveals that prolonging hardware usage saves $X$ dollars in addition to $Y$ tons $CO_{2e}$, it not only strengthens the case for sustainable practice, but also appeals to stakeholders’ bottom lines, potentially influencing AI project planning and policymaking. This type of translation can then be used 
to support the design of targeted policy instruments, including incentives for sustainable AI practices and public procurement standards.


\section{Conclusion}

In the current report, we present the results of an analytical literature review of the current state of AI environmental impact studies, which  synthesizes 61 works on the environmental impacts of AI published between 2019 and 2025 through an eight-stage life cycle lens. We find that existing work is mainly concentrated on training and inference, with other stages of the life cycle much less studied, and that reporting practices vary substantially in metrics, scope, and methodology. These patterns make cross-study comparison difficult and limit the development of more complete and decision-relevant assessments. As a consequence of this analysis, we therefore argue for life cycle-aware reporting, broader environmental metrics, and clearer methodological disclosure as necessary steps toward more rigorous and actionable evaluation of AI's environmental footprint. We also discuss policy implications, notably the importance of collaboration between AI researchers and policymakers and the development of more comprehensive and rigorous metrics for environmental impact assessment, which can then be translated into international standards for regulation and procurement, thereby steering the industry toward more sustainable practices.

\clearpage
\section*{Generative AI Usage Statement}

The authors utilized GPT-5.2 to assist with style editing, grammar correction, the rephrasing of specific sentences to improve clarity, and for brainstorming title options.


\bibliographystyle{ACM-Reference-Format}
\bibliography{bibliography}


\appendix

\section{Overview of Reviewed Papers}
\scriptsize

\begin{longtable}{|p{1.2cm}|p{4cm}|p{0.8cm}|p{3cm}|p{2.5cm}|p{3cm}|}
\hline
Paper number & Paper name & Year & Authors & Venue & life cycle Stage \\ \hline
1 & Energy and Policy Considerations for Deep Learning in NLP~\cite{strubell2019energy} & 2019 & Emma Strubell, Ananya Ganesh, Andrew McCallum & Proceedings of the 57th Annual Meeting of the Association for Computational Linguistics (ACL 2019) & Hyperparameter Experimentation, Model Training, Inference \& Deployment \\
2 & Carbon Emissions and Large Neural Network Training~\cite{patterson2021carbon} & 2021 & David A. Patterson, Joseph Gonzalez, Quoc Le, Chen Liang, Lluis-Miquel Munguia, Daniel Rothchild, David So, Maud Texier, Jeff Dean & arXiv preprint (CoRR) & "Data Center (construction, cooling, etc.)", Model Training \\
3 & Green AI~\cite{schwartz2020green} & 2019 & Roy Schwartz, Jesse Dodge, Noah A. Smith, Oren Etzioni & Communications of the ACM / arXiv preprint & Model Training, Inference \& Deployment \\
4 & Quantifying the Carbon Emissions of Machine Learning~\cite{lacoste2019quantifying} & 2019 & Alexandre Lacoste, Alexandra Luccioni, Victor Schmidt, Thomas Dandres & arXiv preprint & Model Training, "Data Center (construction, cooling, etc.)", Hyperparameter Experimentation \\
5 & Carbontracker: Tracking and Predicting the Carbon Footprint of Training Deep Learning Models~\cite{anthony2020carbontracker} & 2020 & Lasse F. Wolff Anthony, Benjamin Kanding, Raghavendra Selvan & ICML 2020 Workshop on Challenges in Deploying and Monitoring Machine Learning Systems & Model Training, Hyperparameter Experimentation, "Data Center (construction, cooling, etc.)" \\
6 & Life-Cycle Emissions of AI Hardware: A Cradle-To-Grave Approach and Generational Trends~\cite{schneider2025life} & 2025 & Ian Schneider, Hui Xu, Stephan Benecke, David Patterson, Keguo Huang, Parthasarathy Ranganathan, Cooper Elsworth & arXiv preprint (CoRR) & Hardware Manufacturing, "Data Center (construction, cooling, etc.)", Model Training, End-of-Life, Inference \& Deployment \\
7 & Life Cycle Assessment of Artificial Intelligence Applications: Research Gaps and Opportunities~\cite{plociennik2025life} & 2025 & Christiane Plociennik, Ponnapat Watjanatepin, Karel Van Acker, Martin Ruskowski & Procedia CIRP (32nd CIRP Conference on Life Cycle Engineering) & "Dataset gathering, acquisition / generation", Inference \& Deployment, Model Training, Hardware (GPU/CPU life cycle) \\
8 & How Hungry is AI? Benchmarking Energy, Water, and Carbon Footprint of LLM Inference~\cite{jegham2025hungry} & 2025 & Nidhal Jegham, Marwan Abdelatti, Chan Young Koh, Lassad Elmoubarki, Abdeltawab Hendawi & arXiv preprint & Inference \& Deployment, "Data Center (construction, cooling, etc.)" \\
9 & Double-Exponential Increases in Inference Energy: The Cost of the Race for Accuracy~\cite{yang2024double} & 2024 & Zeyu Yang, Karel Adámek, Wesley Armour & arXiv preprint (CoRR) & Inference \& Deployment \\
10 & Reducing the Carbon Impact of Generative AI Inference (today and in 2035)~\cite{chien2023reducing} & 2023 & Andrew A. Chien, Liuzixuan Lin, Hai Nguyen, Varsha Rao, Tristan Sharma, Rajini Wijayawardana & HotCarbon '23 (2nd Workshop on Sustainable Computer Systems) & Inference \& Deployment, "Data Center (construction, cooling, etc.)", Hardware (GPU/CPU life cycle) \\
11 & More than Carbon: Cradle-to-Grave environmental impacts of GenAI training on the Nvidia A100 GPU~\cite{falk2025more} & 2025 & Sophia Falk, David Ekchajzer, Thibault Pirson, Etienne Lees-Perasso, Augustin Wattiez, Lisa Biber-Freudenberger, Sasha Luccioni, Aimee van Wynsberghe & arXiv preprint (CoRR, abs/2509.00093) & Hardware Manufacturing, Hardware (GPU/CPU life cycle), Model Training, End-of-Life \\
12 & The Rising Costs of Training Frontier AI Models~\cite{cottier2024rising} & 2025 & Ben Cottier, Robi Rahman, Loredana Fattorini, Nestor Maslej, David Owen & arXiv preprint (CoRR, abs/2405.21015) & Model Training, Hyperparameter Experimentation, "Data Center (construction, cooling, etc.)" \\
13 & The Hidden Cost of an Image: Quantifying the Energy Consumption of AI Image Generation~\cite{bertazzini2025hidden} & 2025 & Giulia Bertazzini, Chiara Albisani, Daniele Baracchi, Dasara Shullani, Roberto Verdecchia & arXiv preprint (CoRR, abs/2506.17016) & Inference \& Deployment \\
14 & Measuring the Energy Consumption and Efficiency of Deep Neural Networks: An Empirical Analysis and Design Recommendations~\cite{tripp2024measuring} & 2024 & Charles Edison Tripp, Jordan Perr-Sauer, Jamil Gafur, Ambarish Nag, Avi Purkayastha, Sagi Zisman, Erik A. Bensen & arXiv preprint (CoRR, abs/2403.08151) & Model Training, Hyperparameter Experimentation \\
15 & Towards an Energy Consumption Index for Deep Learning Models: A Comparative Analysis of Architectures, GPUs, and Measurement Tools~\cite{aquino2025towards} & 2025 & Sergio Aquino-Brítez, Pablo García-Sánchez, Andrés Ortiz, Diego Aquino-Brítez & Sensors 25(3):846 (MDPI) & Model Training, Inference \& Deployment \\
16 & Benchmarking Energy Efficiency of Large Language Models Using vLLM~\cite{pronk2025benchmarking} & 2025 & K. Pronk, Q. Zhao & arXiv preprint (CoRR, abs/2509.08867) & Inference \& Deployment \\
17 & Towards Sustainable NLP: Insights from Benchmarking Inference Energy in Large Language Models~\cite{poddar2025towards} & 2025 & Soham Poddar, Paramita Koley, Janardan Misra, Niloy Ganguly, Saptarshi Ghosh & Proceedings of the 2025 Conference of the Americas Chapter of the Association for Computational Linguistics: Human Language Technologies (NAACL 2025) & Inference \& Deployment \\
18 & AI and the Net-Zero Journey: Energy Demand, Emissions, and the Potential for Transition~\cite{devarakota2025ai} & 2025 & Pandu Devarakota, Nicolas Tsesmetzis, Faruk O. Alpak, Apurva Gala, Detlef Hohl & arXiv preprint (cs.AI, abs/2507.10750) & Model Training, Inference \& Deployment, "Data Center (construction, cooling, etc.)" \\
19 & Energy Considerations of Large Language Model Inference and Efficiency Optimizations~\cite{fernandez2025energy} & 2025 & Jared Fernandez, Clara Na, Vashisth Tiwari, Yonatan Bisk, Sasha Luccioni, Emma Strubell & Proceedings of the 63rd Annual Meeting of the Association for Computational Linguistics (ACL 2025), Volume 1: Long Papers & Inference \& Deployment \\
20 & Offline Energy-Optimal LLM Serving: Workload-Based Energy Models for LLM Inference on Heterogeneous Systems~\cite{wilkins2024offline} & 2024 & Grant Wilkins, Srinivasan Keshav, Richard Mortier & Proceedings of the 3rd ACM HotCarbon Workshop on Sustainable Computer Systems (HotCarbon '24) & Inference \& Deployment \\ 
21 &
The ML.ENERGY Benchmark: Toward Automated Inference Energy Measurement and Optimization~\cite{chung2025ml} &
2025 &
Jae-Won Chung; Jiachen Liu; Jeff J. Ma; Ruofan Wu; Oh Jun Kweon; Yuxuan Xia; Zhiyu Wu; Mosharaf Chowdhury &
NeurIPS 2025, Datasets and Benchmarks Track (arXiv:2505.06371) &
Inference \& Deployment \\
22 &
Estimating the Carbon Footprint of BLOOM, a 176B Parameter Language Model~\cite{luccioni2023estimating} &
2023 &
Alexandra Sasha Luccioni; Sylvain Viguier; Anne-Laure Ligozat &
Journal of Machine Learning Research (JMLR), 24:1--15 &
Model Training; Inference \& Deployment; Hardware Manufacturing; Data Center (construction, cooling, etc.) \\
23 &
Measuring and Improving the Energy Efficiency of Large Language Models Inference~\cite{argerich2024measuring} &
2024 &
Mauricio Fadel Argerich; Marta Patino-Martinez &
IEEE Access, 12:80194--80207 &
Inference \& Deployment \\
24 &
How to estimate carbon footprint when training deep learning models? A guide and review~\cite{bouza2023estimate} &
2023 &
Lucia Bouza Heguerte; Aurelie Bugeau; Loic Lannelongue &
Environmental Research Communications, 5(11):115014 &
Model Training \\
25 &
Measuring the environmental impact of delivering AI at Google Scale~\cite{elsworth2025measuring} &
2025 &
Cooper Elsworth; Keguo Huang; David Patterson; Ian Schneider; Robert Sedivy; Savannah Goodman; Ben Townsend; \\
   & & & Parthasarathy Ranganathan; Jeff Dean; Amin Vahdat; Ben Gomes; James Manyika &
Google Research technical report / whitepaper &
Inference \& Deployment; Hardware (GPU/CPU lifecycle); Data Center (construction, cooling, etc.); Model Training; Dataset gathering, acquisition / generation \\
26 &
From Words to Watts: Benchmarking the Energy Costs of LLM Inference~\cite{samsi2023words} &
2023 &
Siddharth Samsi; Dan Zhao; Joseph McDonald; Baolin Li; Adam Michaleas; Michael Jones; \\
   & & & William Bergeron; Jeremy Kepner; Devesh Tiwari; Vijay Gadepally &
arXiv preprint arXiv:2310.03003 &
Inference \& Deployment \\
27 &
Computing Within Limits: An Empirical Study of Energy Consumption in ML Training and Inference~\cite{mavromatis2024computing} &
2024 &
Ioannis Mavromatis; Kostas Katsaros; Aftab Khan &
ARISDE 2024: 1st International Workshop on Artificial Intelligence for Sustainable Development &
Model Training \\
28 &
The Carbon Footprint of Machine Learning Training Will Plateau, Then Shrink~\cite{patterson2022carbon} &
2022 &
David Patterson; Joseph Gonzalez; Urs Holzle; Quoc Le; Chen Liang; Lluis-Miquel Munguia; \\
   & & & Daniel Rothchild; David So; Maud Texier; Jeff Dean &
IEEE Computer (journal; also arXiv:2204.05149) &
Inference \& Deployment \\
29 &
Counting Carbon: A Survey of Factors Influencing the Emissions of Machine Learning~\cite{luccioni2023counting} &
2023 &
Alexandra Sasha Luccioni; Alex Hernandez-Garcia &
arXiv preprint arXiv:2302.08476 &
Model Training \\
30 &
Energy-Aware Machine Learning Models---A Review of Recent Techniques and Perspectives~\cite{rozycki2025energy} &
2025 &
Rafal Rozycki; Dorota Agnieszka Solarska; Grzegorz Waligora &
Energies (MDPI), 18(11):2810 &
Model Training \\
31 & Method and evaluations of the effective gain of artificial intelligence models for reducing CO$_2$ emissions~\cite{delanoe2023method} & 2023 & Paul Delanoe, Dieudonne Tchuente, Guillaume Colin & Journal of Environmental Management & Model Training, Inference \& Deployment, End-of-Life \\
32 & Comparative Study of GPU Performance and Energy Efficiency Across Generational Architectures: A Systematic Literature~\cite{muhammad2024comparative} & 2024 & Risang Faiz Muhammad, Muhammad Edo Syahputra & 2024 IEEE International Conference on Control \& Automation, Electronics, Robotics, Internet of Things, and Artificial Intelligence (CERIA 2024) &
Hardware (GPU/CPU Lifecycle) \\
33 & Single-Node Power Demand During AI Training: Measurements on an 8-GPU NVIDIA H100 System~\cite{latif2025single} & 2025 & Imran Latif, Alex C. Newkirk, Matthew R. Carbone, Arslan Munir, Yuewei Lin, Jonathan G. Koomey, Xi Yu, Zhihua Dong & IEEE Access & Model Training \\
34 & Sustainable Supercomputing for AI: GPU Power Capping at HPC Scale~\cite{zhao2023sustainable} & 2023 & Dan Zhao, Siddharth Samsi, Joseph McDonald, Baolin Li, David Bestor, Michael Jones, Devesh Tiwari, Vijay Gadepally & ACM Symposium on Cloud Computing (SoCC '23) & Hardware (GPU/CPU Lifecycle) \\
35 & The Energy Cost of Artificial Intelligence Lifecycle in Communication Networks~\cite{chou2024energy} & 2025 & Shih-Kai Chou, Jernej Hribar, Vid Hanzel, Mihael Mohorcic, Carolina Fortuna & arXiv preprint (arXiv:2408.00540) & "Dataset gathering, acquisition / generation" \\
36 & Power Hungry Processing: Watts Driving the Cost of AI Deployment?~\cite{luccioni2024power} & 2024 & Alexandra Sasha Luccioni, Yacine Jernite, Emma Strubell & arXiv preprint (arXiv:2311.16863) & Inference \& Deployment \\
37 & BoaviztAPI: A Bottom-Up Model to Assess the Environmental Impacts of Cloud Services~\cite{simon2024boaviztapi} & 2025 & Thibault Simon, David Ekchajzer, Adrien Berthelot, Eric Fourboul, Samuel Rince, Romain Rouvoy & ACM SIGEnergy Energy Informatics Review (EIR), 4(5):84--90 & Hardware (GPU/CPU Lifecycle), "Data Center (construction, cooling, etc.)" \\
38 & Ground-Truthing AI Energy Consumption: Validating CodeCarbon Against External Measurements~\cite{fischer2025ground} & 2025 & Raphael Fischer & arXiv preprint (arXiv:2509.22092) & Inference \& Deployment, Hyperparameter Experimentation \\
39 & A framework for measuring the training efficiency of a neural architecture\cite{cueto2024framework} & 2024 & Eduardo Cueto-Mendoza, John D. Kelleher & Artificial Intelligence Review, 57:349 & Model Training, Hardware (GPU/CPU Lifecycle) \\
40 & Trends in AI inference energy consumption: Beyond the performance-vs-parameter laws of deep learning~\cite{desislavov2023trends} & 2023 & Radosvet Desislavov, Fernando Martinez-Plumed, Jose Hernandez-Orallo & Sustainable Computing: Informatics and Systems, 38:100857 & Inference \& Deployment \\
41 & The water use of data center workloads: A review and assessment of key determinants~\cite{lei2025water} & 2025 & Nuoa Lei, Jun Lu, Arman Shehabi, Eric Masanet & Resources, Conservation \& Recycling & "Data Center (construction, cooling, etc.)", Model Training, Inference \& Deployment, End-of-Life \\
42 & Sustainable AI infrastructure: A scenario-based forecast of water footprint under uncertainty~\cite{herrera2025sustainable} & 2025 & Manuel Herrera, Xiang Xie, Andrea Menapace, Ariele Zanfei, Bruno M. Brentan & Journal of Cleaner Production & "Data Center (construction, cooling, etc.)" \\
43 & Digital \& environment : How to evaluate server manufacturing footprint, beyond greenhouse gas emissions?~\cite{lorenzini2021digital} & 2021 & Romain Lorenzini & Boavizta blog (Digital \& environment) & Hardware Manufacturing, End-of-Life, "Data Center (construction, cooling, etc.)" \\
44 & Reconciling the contrasting narratives on the environmental impact of large language models~\cite{ren2024reconciling} & 2024 & Shaolei Ren, Bill Tomlinson, Rebecca W. Black, Andrew W. Torrance & Scientific Reports (Nature) & Inference \& Deployment, "Data Center (construction, cooling, etc.)", Model Training \\
45 & Misinformation by Omission: The Need for More Environmental Transparency in AI~\cite{luccioni2025misinformation} & 2025 & Sasha Luccioni, Boris Gamazaychikov, Theo Alves da Costa, Emma Strubell & arXiv preprint arXiv:2506.15572 & Inference \& Deployment, "Data Center (construction, cooling, etc.)", Model Training, Hardware Manufacturing \\
46 & Exploring energy consumption of AI frameworks on a 64-core RV64 Server CPU~\cite{malenza2025exploring} & 2025 & Giulio Malenza, Francesco Targa, Adriano Marques Garcia, Marco Aldinucci, Robert Birke & arXiv preprint arXiv:2504.03774 & Inference \& Deployment \\
47 & EcoLogits: Evaluating the Environmental Impacts of Generative AI~\cite{rince2025ecologits} & 2025 & Samuel Rincé, Adrien Banse & Journal of Open Source Software (JOSS) & Hardware (GPU/CPU Lifecycle), Inference \& Deployment \\
48 & Evaluating the Environmental Impact of Language Models with Life Cycle Assessment~\cite{fernandezevaluating} & 2025 & Jared Fernandez, Clara Na, Yonatan Bisk, Emma Strubell & Climate Change AI Workshop at ICLR 2025 (Proposals Track) & Hardware Manufacturing, Model Training, Inference \& Deployment \\
49 & Holistically Evaluating the Environmental Impact of Creating Language Models~\cite{morrison2025holistically} & 2025 & Jacob Morrison, Clara Na, Jared Fernandez, Tim Dettmers, Emma Strubell, Jesse Dodge & International Conference on Learning Representations (ICLR 2025) & Hyperparameter Experimentation, Hardware (GPU/CPU Lifecycle), Model Training, Inference \& Deployment \\
50 & Making AI Less ``Thirsty'': Uncovering and Addressing the Secret Water Footprint of AI Models~\cite{li2025making} & 2023 & Pengfei Li, Jianyi Yang, Mohammad A. Islam, Shaolei Ren & arXiv preprint arXiv:2304.03271 / Communications of the ACM & "Data Center (construction, cooling, etc.)", Inference \& Deployment, Model Training, Supply-chain manufacturing \\
51 & Carbon Footprint of AI Data Centers: A Life Cycle Approach~\cite{d2024carbon} & 2025 & Alexandre d'Orgeval, Edi Assoumou, Valentina Sessa, Ilknur Colak, Stuart Sheehan, Quentin Avenas & Energy Proceedings, 16th International Conference on Applied Energy (ICAE 2024) & "Data Center (construction, cooling, etc.)" \\
52 & Towards the Systematic Reporting of the Energy and Carbon Footprints of Machine Learning~\cite{henderson2020towards} & 2020 & Peter Henderson, Jieru Hu, Joshua Romoff, Emma Brunskill, Dan Jurafsky, Joelle Pineau & Journal of Machine Learning Research (JMLR) & Hyperparameter Experimentation, Model Training \\
53 & Green and intelligent: the role of AI in the climate transition~\cite{stern2025green} & 2025 & Nicholas Stern, Mattia Romani, Roberta Pierfederici, Manuel Braun, Daniel Barraclough, Shajeeshan Lingeswaran, Elizabeth Weirich-Benet, Niklas Niemann & npj Climate Action & "Data Center (construction, cooling, etc.)" \\
54 & E-waste challenges of generative artificial intelligence~\cite{wang2024waste} & 2024 & Peng Wang, Ling-Yu Zhang, Asaf Tzachor, Wei-Qiang Chen & Nature Computational Science & Hardware (GPU/CPU Lifecycle), End-of-Life \\
55 & Frameworks for the application of machine learning in life cycle assessment for process modeling~\cite{martinez2024frameworks} & 2024 & Nicolás Martínez-Ramón, Fernando Calvo-Rodríguez, Diego Iribarren, Javier Dufour & Cleaner Environmental Systems & None \\
56 & Life cycle assessment of a climate-friendly data center cooling device~\cite{isler2023life} & 2023 & Asli Isler-Kaya, Filiz Karaosmanoglu & Energy and Buildings & "Data Center (construction, cooling, etc.)" \\\\
57 & Using life cycle assessment to drive innovation for sustainable cool clouds~\cite{alissa2025using} & 2025 & Husam Alissa, Teresa Nick, Ashish Raniwala, Alberto Arribas Herranz, Kali Frost, Ioannis Manousakis, Kari Lio, Brijesh Warrier & Nature & Inference \& Deployment, "Data Center (construction, cooling, etc.)" \\
58 & Unraveling the Hidden Environmental Impacts of AI Solutions for Environment Life Cycle Assessment of AI Solutions~\cite{ligozat2022unraveling} & 2022 & Anne-Laure Ligozat, Julien Lefevre, Aurélie Bugeau, Jacques Combaz &
Sustainability & Dataset gathering, acquisition / generation; Model Training; Inference \& Deployment; Hardware Manufacturing; Hardware (GPU/CPU Lifecycle); Data Center (construction, cooling, etc.); End-of-Life \\
59 & Estimating the environmental impact of Generative-AI services using an LCA-based methodology~\cite{berthelot2024estimating} &
2024 & Adrien Berthelot, Eddy Caron, Mathilde Jay, Laurent Lefèvre & Procedia CIRP (31st Conference on Life Cycle Engineering) & Model Training; Inference \& Deployment; Hardware Manufacturing; Hardware (GPU/CPU Lifecycle); Data Center (construction, cooling, etc.); End-of-Life \\
60 & Sustainable AI: Environmental Implications, Challenges and Opportunities~\cite{wu2022sustainable} &
2022 & Carole-Jean Wu, Ramya Raghavendra, Udit Gupta, Bilge Acun, Newsha Ardalani, Kiwan Maeng, Gloria Chang, Fiona Aga Behram, James Huang, Charles Bai, Michael Gschwind, Anurag Gupta, Myle Ott, Anastasia Melnikov, Salvatore Candido, David Brooks, Geeta Chauhan, Benjamin Lee, Hsien-Hsin S. Lee, Bugra Akyildiz, Maximilian Balandat, Joe Spisak, Ravi Jain, Mike Rabbat, Kim Hazelwood &
Proceedings of the 5th Conference on Machine Learning and Systems (MLSys 2022) &
"Dataset gathering, acquisition / generation"; Hyperparameter Experimentation; Model Training; Inference \& Deployment; Hardware Manufacturing; Hardware (GPU/CPU Lifecycle); "Data Center (construction, cooling, etc.)" \\

61 & Beyond Counting Carbon: AI
Environmental Assessments
Struggle to Inform Net Impact
Decisions~\cite{cook2025beyond}& 2025 & Cook, Jacob, Walton, Berthelot, Hussain, Schien & ETH Research Collection & Model Training; Inference \& Hardware Manufacturing; Infrastructure Construction; Model Training; Deployment \& Inference \\

\label{app:paper-table}
\end{longtable}

\end{document}